\documentclass[prd,aps,11pt,
nofootinbib,tightenlines,superscriptaddress,preprintnumbers]{revtex4}

\def\lsim{\raise0.3ex\hbox{$<$\kern-0.75em\raise-1.1ex\hbox{$\sim$}}}
\def\gsim{\raise0.3ex\hbox{$>$\kern-0.75em\raise-1.1ex\hbox{$\sim$}}}

\usepackage{graphicx}

\newcommand{\beq}{\begin{equation}}
\newcommand{\eeq}{\end{equation}}
\newcommand{\bqa}{\begin{eqnarray}}
\newcommand{\eqa}{\end{eqnarray}}

\begin{document}

\title{Causal Viscous Hydrodynamics for Central Heavy-Ion Collisions II:
Meson Spectra and HBT Radii}

\preprint{INT PUB 07-03}

\author{Paul Romatschke}
\affiliation{Institute for Nuclear Theory, University of Washington,
Box 351550, Seattle WA, 98195, USA}
\date{\today}

\begin{abstract} 
Causal viscous hydrodynamic fits to experimental data for 
pion and kaon transverse momentum spectra from central Au+Au 
collisions at $\sqrt{s_{NN}}=200$ GeV are presented. 
Starting the hydrodynamic evolution at 1 fm/c and using 
small values for the relaxation time, reasonable fits
up to moderate ratios $\eta/s\simeq0.4$ can be obtained.
It is found that a percentage of roughly $50\, \eta/s$ 
to $75\, \eta/s$ of the final
meson multiplicity is due to viscous entropy production.
Finally, it is shown that with increasing viscosity, the ratio of
HBT radii $R_{\rm out}/R_{\rm side}$ approaches 
and eventually matches the experimental data.
\end{abstract}

\maketitle

\section{Introduction}

Extracting a value for the shear viscosity $\eta$ out of data from
the ongoing heavy-ion collision program at the 
Relativistic Heavy-Ion Collider (RHIC) is a difficult but maybe rewarding 
issue. It is difficult because the simplest hydrodynamic theory that
includes viscosity, Navier-Stokes theory, is known to have 
problems with causality and instabilities in the relativistic case
\cite{Hisc}. In order to repair these problems, so-called second-order 
theories have been put forward by Israel, Stewart \cite{IS} and Liu,
M\"uller, Ruggeri \cite{LMR}. Unfortunately, this formulation 
introduces at least one
other a priori unconstrained parameter, a relaxation time, into the 
hydrodynamic framework. Also, the resulting hydrodynamic equations
are quite complicated, and not easily implemented using
existing numerical 
schemes\footnote{In the context of heavy-ion collisions, solutions for
causal viscous hydrodynamic theories without
transverse or longitudinal dynamics (only ``Bjorken flow'')
have been obtained in 
Refs.~\cite{Muronga:2001zk,Muronga:2003ta,Baier:2006um,Koide:2006ef}
while results including transverse flow can be found
in Refs.~\cite{Muronga:2004sf,Chaudhuri:2005ea,Baier:2006gy}.
No results including elliptic flow exist up to date.}.

Extracting $\eta$ from experiment may be rewarding, however,
as results for $\eta$ 
from weak-coupling calculations for QCD \cite{Trans1,Trans2} 
and strong-coupling calculations
for ${\mathcal N}=4$ SYM \cite{Kovtun:2004de,Janik:2006ft} 
differ by an order of magnitude. Thus the value extracted from
experiment might offer a clue whether the quark-gluon plasma
at RHIC is described better by weak or strong coupling 
techniques\footnote{An anomalous viscosity value from turbulent magnetic fields 
\cite{Asakawa:2006tc}
due to plasma instabilities \cite{Mrowczynski:2005ki}
could blur this naive picture, however.}, a question
currently hotly debated in the physics community.

On the practical level, determining $\eta$ (or more precisely
the ratio of shear viscosity over entropy density, $\eta/s$) from
experiment translates to finding the value of $\eta/s$ for which
a viscous hydrodynamic model fits experimental data best.
Unfortunately, there is a lot of freedom in any hydrodynamic model
calculation, mostly because of the poorly constrained initial conditions,
but also in the ``final'' or freeze-out conditions (see below).
It turns out that results for central collisions alone are not sufficient 
to fix all the free parameters, and as a consequence only
upper bounds on the ratio of shear viscosity over entropy 
can be given in this work. 
Moreover, it should be noted
that besides shear viscosity also bulk viscosity
and heat-conductivity will affect hydrodynamic model fits to experimental
data\footnote{Bulk viscosity, however, has recently been
calculated in QCD using weak-coupling techniques and found to be negligible
compared to shear viscosity \cite{Arnold:2006fz}.}. 
Algorithms on how to include these have been suggested in
\cite{Heinz:2005bw,Muronga:2006zw}. In order to keep complexity
to a minimum, in the following only the effect of shear viscosity
is included.
This work is organized as follows: in section \ref{sec:two}, 
the setup of causal viscous hydrodynamics as well as the 
initial conditions and equation of state for heavy-ion collisions
are briefly reviewed. In section \ref{sec:three}, the 
equations for calculating particle spectra and HBT radii in
viscous hydrodynamics are given and results are compared
to experimental data in section \ref{sec:four}. Section
\ref{sec:five} contains a summary and the conclusions.

\section{Setup}
\label{sec:two}
Including only the effects of shear viscosity, the causal
viscous hydrodynamic equations 
used in the following are given by \cite{Baier:2006um}
\bqa
(\epsilon+p)D u^\mu&=&\nabla^\mu p-\Delta^\mu_{\nu} \nabla_\sigma
\Pi^{\nu \sigma}+\Pi^{\mu \nu} D u_\nu\, ,
\label{2.1}\\
D \epsilon &=& - (\epsilon+p) \nabla_\mu u^\mu+\frac{1}{2}\Pi^{\mu \nu}
\langle\nabla_\nu u_\mu\rangle\, ,
\label{2.2}\\
\tau_{\Pi} \Delta^\mu_\alpha \Delta^\nu_\beta D \Pi^{\alpha \beta}
+\Pi^{\mu \nu}&=&\eta  \langle\nabla^\mu u^\nu\rangle
- 2 \tau_{\Pi} \Pi^{\alpha (\mu}\omega^{\nu)}_{\ \alpha}\, ,
\label{baseq}
\eqa
where $\epsilon,p$ are the energy density and pressure, respectively.
The flow four-velocity $u^\mu$ obeys $u_\mu u^\mu=1$ and
$\Pi^{\mu \nu}$ is the shear tensor that fulfills
$u_\mu \Pi^{\mu \nu}=0=\Pi^\mu_\mu$ and characterizes 
the deviations due to viscosity in the energy momentum tensor,
\beq
T^{\mu \nu}=(\epsilon+p)u^\mu u^\nu-p g^{\mu \nu}+\Pi^{\mu \nu}.
\label{Tdef}
\eeq 
The remaining definitions are
\bqa
&d_\mu u^\nu \equiv \partial_\mu u^\nu+\Gamma_{\alpha \mu}^{\nu}
u^\alpha,\qquad
D\equiv u_\mu d^\mu,\qquad
\nabla^\mu \equiv \Delta^{\mu \nu} d_\nu,&
\nonumber\\
&\Delta^{\mu\nu}\equiv g^{\mu \nu}-u^\mu u^\nu, \qquad
\omega^{\mu \nu}=\Delta^{\mu \alpha} \Delta^{\nu \beta}
\frac{1}{2}\left(d_\beta u_\alpha-d_\alpha u_\beta\right),&\nonumber\\
&\langle A_\mu B_\nu\rangle \equiv  A_\mu B_\nu+A_\nu B_\mu
-\frac{2}{3} \Delta_{\mu \nu} A_\alpha B^\alpha,\qquad
(A_\mu,B_\nu) \equiv 
\frac{1}{2}\left(A_\mu B_\nu+A_\nu B_\mu\right),&
\label{symbdef}
\eqa
where $\Gamma_{\alpha \mu}^\nu$ are the Christoffel symbols.
The parameter $\tau_\Pi$ is a relaxation time that in weakly 
coupled QCD can be related to $\eta$ and the pressure as 
\cite{Muronga:2003ta,Baier:2006um}
$\tau_\Pi \simeq\frac{3\eta}{2 p}$, which translates to
$\tau_\Pi \simeq\frac{\eta}{s}\frac{6}{T}$. 
To test for the dependence of the results
on $\tau_\Pi$, in the following also a 
somewhat smaller value $\tau_\Pi \simeq\frac{\eta}{s}\frac{1.5}{T}$ 
will be used\footnote{
Once a non-trivial equation of state is used, final results will
differ depending to whether one defines $\tau_\Pi=\frac{3 \eta}{2 p}$
or $\tau_\Pi=\frac{\eta}{s} \frac{6}{T}$. In this work, 
the latter definition is adopted,
but final results seem to differ only slightly when implementing the other
choice.},
which can be argued for independently \cite{prep}.
Note that formally one recovers the relativistic Navier-Stokes
equations from Eq.~(1,2,3) in the limit $\tau_\Pi\rightarrow 0$.

The algorithm to solve the above equations including 
several tests was outlined
in detail in Ref.~\cite{Baier:2006gy} and is not repeated here
for brevity. 
In this work, the equations are solved on a lattice with $512$ sites
and a lattice spacing of $0.25\, {\rm GeV}^{-1}$.

\subsection{Initial Conditions and Equation of State}

The energy density at the hydrodynamic initialization time 
$\tau_0$ is assumed to be parameterized by the number density
of wounded nucleons in a Glauber model \cite{Kolb:2001qz},
\bqa
\epsilon(\tau_0,r)&=&{\rm const}\times n_{\rm WN}(r)
\label{edinit}\\
n_{\rm WN}(r)&=&2\, T_A(r)\left[1-\left(1-
\frac{\sigma T_A(r)}{A}\right)^A\right]\nonumber\\
T_A(r)&=&\int_{-\infty}^{\infty} dz \frac{\rho_0}{1
+\exp[(\sqrt{r^2+z^2}-R_0)/\chi]}\nonumber,
\eqa
where $\rho_0$ is such that $2\pi \int r  dr\ T_A(r)=A$.
For gold nuclei, $A=197$, $R_0=6.4$ fm, $\chi=0.54$ fm and for numerical
reasons $\left(1-\frac{\sigma T_A(r)}{A}\right)^A$ is replaced by
an exponential. The nucleon-nucleon cross section at $\sqrt{s_{NN}}=200$ 
GeV is assumed to be given by $\sigma=40$ mb.
Other parameterization of the initial energy density (e.g.
scaling by the number of binary collisions), 
in general result in stronger radial gradients \cite{Kolb:2001qz}
and consequently faster buildup of transverse flow.
Since viscosity in some sense mimics the presence of transverse flow,
the least restrictive initial condition and therefore the most 
conservative bound on viscosity will come from the choice of 
Eq.~(\ref{edinit}).
In the same spirit, the initial value of $\Pi^{\mu\nu}$ is chosen
to be zero everywhere (see also the discussion in Ref.~\cite{Baier:2006gy}).

The constant in Eq.~(\ref{edinit}) is chosen such that the central
energy density $\epsilon(r=0)$ corresponds to a predefined
starting temperature $T_0$ via the equation of state.
Since lattice QCD seems to rule out a first or second order 
phase transition \cite{fodor}, the semi-realistic equation of state of Laine 
and Schr\"oder \cite{Laine:2006cp} is used in the following.
This equation of state is calculated from a hadron resonance gas
at low temperatures and high-order weak-coupling QCD result at high
temperatures with a cross-over transition near $T_c\sim 175$ MeV
(for brevity, the reader is referred to \cite{Laine:2006cp} for details).

\section{Particle Spectra and HBT Radii in Viscous Hydrodynamics}
\label{sec:three}
\subsection{Particle Spectra}

In order to convert hydrodynamic quantities into experimentally interesting
observables, the standard method of choice is the Cooper-Frye
freeze-out prescription \cite{CooperFrye}. Assuming isothermal
freeze-out at the temperature $T_f$ defines a freeze-out surface
which is characterized \cite{Rischke:1996em} 
by its normal vector $d\Sigma^\mu$,
\beq
\left(d\Sigma^t,d\Sigma^x,d\Sigma^y,d\Sigma^z\right)=
\left(\cosh(\eta),\cos(\phi) \frac{d\tau(\zeta)}{d \zeta},
\sin(\phi) \frac{d\tau(\zeta)}{d \zeta}, \sinh(\eta)\right) 
r(\zeta) \tau(\zeta) \frac{dr(\zeta)}{d \zeta} d\zeta d\phi d\phi.
\eeq
The surface is parameterized by $\zeta\, \epsilon\, [0,1]$ such that 
$\tau(0)=\tau_0$ and $\tau(1)$ corresponding to the time when the
last fluid element has cooled down to the temperature $T_f$.
The single particle spectra 
for a particle with four momentum $p^\mu=(E,{\bf p})$ and degeneracy
$d$ are then calculated as
\beq
E\frac{d^3 N}{d^3 {\bf p}}\equiv \frac{d}{(2 \pi)^3}
\int p_\mu d\Sigma^\mu f\left(\frac{p_\mu u^\mu}{T}\right),
\label{OPdef}
\eeq
where it is reminded that the viscous distribution function $f$
is related to the ideal distribution 
$f_0(x)=[\exp(x)\pm 1]^{-1}$ and the shear tensor
$\Pi^{\mu \nu}$ as \cite{Teaney, Baier:2006um} 
\beq
f\left(\frac{p_\mu u^\mu}{T}\right)=f_0\left(\frac{p_\mu u^\mu}{T}\right)
\left[1\mp f_0\left(\frac{p_\mu u^\mu}{T}\right)\right]
\left[1+\frac{p_\mu p_\nu \Pi^{\mu \nu}}{2 T^2 (\epsilon+p)}\right],
\label{fequation}
\eeq
where $\pm$ applies for fermions and bosons, respectively. For 
simplicity it is, however, convenient to approximate
$f_0(x)\simeq \exp{(-x)}$ which does not seem to affect final
results too much. In this case, one consequently also has to replace
the expression $[1\mp f(x)]$ in Eq.~(\ref{fequation}) by $1$
and all but one integral in Eq.~(\ref{OPdef}) can be evaluated 
analytically to give
\cite{Baier:2006gy}
\bqa
E \frac{d^3\, N}{d^3p} &=&
-\frac{2 d}{(2\pi)^2} \int_0^1 d \zeta\  r(\zeta)\  \tau(\zeta)
\left\{
\left[\frac{dr}{d\, \zeta}\,
m_\perp I_0(u^r p_\perp/T) K_1(u^\tau m_\perp/T)\right.\right.\nonumber\\
&&\left.\left.\hspace*{6cm}-\frac{d\tau}{d\,\zeta} \, p_\perp I_1(u^r p_\perp/T) K_0(u^\tau m_\perp/T
)\right]\right.\nonumber\\
&&\left.+\frac{2 p_\perp m_\perp v \Pi^r_r}{4T^2 (\epsilon+p)} 
\left[\frac{dr}{d\, \zeta}\,m_\perp(K_0+K_2) I_1-
\frac{d\tau}{d\, \zeta} p_\perp K_1 (I_2+I_0)\right]
\right.\nonumber\\
&&\left.
-\frac{p_\perp^2 \Pi^r_r}{4T^2 (\epsilon+p)} 
\left[\frac{dr}{d\, \zeta}\,2 m_\perp K_1 I_2-
\frac{d\tau}{d\,\zeta} \, p_\perp
K_0 (I_3+I_1)\right]\right.\nonumber\\
&&\left.
-\frac{v^2 m_\perp^2 \Pi^r_r}{4T^2 (\epsilon+p)} 
\left[\frac{dr}{d\, \zeta}\,\frac{1}{2}m_\perp(3 K_1+K_3) I_0-
\frac{d\tau}{d\,\zeta}\, p_\perp (K_0+K_2) I_1\right]\right.\nonumber\\
&&\left.
-\frac{v^2 p_\perp^2 \Pi^r_r}{4T^2 (\epsilon+p)} 
\left[\frac{dr}{d\, \zeta}\,m_\perp K_1 (I_0-I_2)-
\frac{d\tau}{d\,\zeta}\, p_\perp K_0 \frac{1}{2} (I_1-I_3)\right]\right.\nonumber\\
&&\left.
+\frac{p_\perp^2 \Pi^\eta_\eta}{4T^2 (\epsilon+p)}  
\left[\frac{dr}{d\, \zeta}\,m_\perp K_1 (I_0-I_2)-
\frac{d\tau}{d\,\zeta}\, p_\perp K_0 \frac{1}{2} (I_1-I_3)\right]\right.\nonumber\\
&&\left.
-\frac{m_\perp^2 \Pi^\eta_\eta}{4T^2 (\epsilon+p)}
\left[\frac{dr}{d\, \zeta}\,m_\perp \frac{1}{2} (K_3-K_1) I_0
-\frac{d\tau}{d\,\zeta}\, p_\perp (K_2-K_0) I_1\right]
\right\},
\label{d3Nres}
\eqa
where $I_n(x)$ and $K_n(y)$ are modified Bessel functions that have arguments
$x=u^r p_\perp/T$ and $y=u^\tau m_\perp/T$ as denoted in the first
two lines.

\subsection{HBT Radii}

Given two identical particles with four momenta $p_1^\mu$ and 
$p_2^\mu$,
respectively, the coincidence probability of measuring these two 
particles in a single event divided by the probability of the
particles being uncorrelated defines the two-particle correlation function
$C(p_1^\mu,p_2^\mu)$. Rewriting the correlation function
as a function of the momentum difference $q^\mu=p_1^\mu-p_2^\mu$
and average momentum $K^\mu=\frac{1}{2}\left(p_1^\mu+p_2^\mu\right)$
and assuming chaoticity and large size of the emitting particle source,
the correlation function may be written as \cite{Schlei:1992jj}
\beq
C(p_1^\mu,p_2^\mu)=1+\frac{E_1 E_2\frac{d^6 N}{d^3 {\bf p_1} d^3 {\bf p_2}}}
{E_1\frac{d^3 N}{d^3 {\bf p_1}} E_2\frac{d^3 N}{d^3 {\bf p_2}}},
\label{Chydrodef}
\eeq
where $E_1\frac{d^3 N}{d^3 {\bf p_1}}$, $E_2\frac{d^3 N}{d^3 {\bf p_2}}$
are the single particle spectra defined earlier and 
\beq
E_1 E_2\frac{d^6 N}{d^3 {\bf p_1} d^3 {\bf p_2}}=\left|
\frac{d}{(2\pi)^3} \int_{\Sigma} d\Sigma_\mu K^\mu 
\exp{\left[i \Sigma_\mu q^\mu\right]}f\left( K_\mu u^\mu\right)
\right|^2.
\label{d6Ndef}
\eeq
If furthermore boost-invariance and rotational symmetry around the 
longitudinal axis is imposed, one can choose the average transverse 
momentum as ${\bf k}=(k^x,k^y,k^z)=(k,0,0)$ and decompose ${\bf q}$ into so-called
``out'', ``side'' and ``long'' components, ${\bf q_{\rm out}}=(q_{\rm out},0,0)$,
${\bf q_{\rm side}}=(0,q_{\rm side},0)$, ${\bf q_{\rm long}}=(0,0,q_{\rm long})$,
respectively. The correlation function $C(p_1^\mu,p_2^\mu)$ for these components 
then also can be used to define the three HBT-radii via the Bertsch-Pratt 
parameterization \cite{Pratt:1986ev,Bertsch:1989vn}
\beq
C({\bf q},{\bf k})\simeq 1+ \exp{\left[-R^2_{\rm out}(k) q_{\rm out}^2 
-R^2_{\rm side}(k) q_{\rm side}^2
-R^2_{\rm long}(k) q_{\rm long}^2\right]}.
\eeq
Even though the correlation function $C$ obtained through Eq.~(\ref{Chydrodef}) 
will in general not have the above Gaussian form, for simplicity
the HBT radii are determined  as $R=\sqrt{\ln{2}}/q^*$ where
$C(q^*,k)=1.5$, as proposed in Ref.~\cite{Muronga:2004sf} (see also
Ref.\cite{Rischke:1996em}).

With the freeze-out surface $\Sigma^\mu$ taking the form \cite{Rischke:1996em}
\beq
\left(\Sigma^t,\Sigma^x,\Sigma^y,\Sigma^z\right)=
\left(\tau(\zeta) \cosh(\eta), r(\zeta) \cos(\phi), r(\zeta) \sin(\phi), \tau(\zeta) \sinh(\eta)\right)
\eeq
one can again do some of the integrals in Eq.~(\ref{d6Ndef}) analytically. Specifically, it is found that
for $R_{\rm long}$, the result for $\frac{d}{(2\pi)^3} \int_{\Sigma} d\Sigma_\mu K^\mu 
\exp{\left[i \Sigma_\mu q^\mu\right]}f\left( K_\mu u^\mu\right)$ becomes an expression similar to
Eq.~(\ref{d3Nres}), but with all Bessel K functions replaced by
\beq
K_n(x) \rightarrow \hat{K}_n \equiv \int_0^\infty \cosh{(n\, \eta)} \cos{[\tau q_{\rm long} \sinh{\eta}]} \exp{[-x \cosh{\eta}]}.
\eeq
For $R_{\rm side}$, one has to replace all Bessel I functions in Eq.~(\ref{d3Nres}) by
\beq
I_n(x) \rightarrow \hat{I}_n \equiv \int_0^\pi \frac{d\phi}{\pi} \cos{(n\, \phi)} \cos{[r q_{\rm side} \sin(\phi)]}\exp{[x \cos(\phi)]}
\eeq
and finally, for $R_{\rm out}$, the arguments of the Bessel I and K functions have to be replaced by
\beq
x\rightarrow \hat{x}\equiv \frac{u^r p_\perp}{T}-i \tau (E_1-E_2), \qquad y\rightarrow \hat{y} \equiv 
\frac{u^\tau m_\perp}{T}-i r q_{\rm out}.
\eeq
(Note that in this case the modulus in Eq.~(\ref{d6Ndef}) is important.)
Parts of these simple relations have been found in \cite{Rischke:1996em,Muronga:2004sf}.

\section{Results}
\label{sec:four}

\subsection{Meson Spectra}

Spectra of pions and kaons have contributions from two sources: 
firstly, there are direct contributions for these particles at freeze-out, 
which are calculated using Eq.~(\ref{d3Nres}). Secondly, there are
contributions from unstable hadrons and hadron resonances that 
can decay into pions and kaons. The spectra of these unstable 
particles with masses up to 2 GeV 
are also calculated at freeze-out via Eq.~(\ref{d3Nres})
and then all possible two- and three-body decays that contribute to the
stable particles of interest are taken into account 
using the decay routine from the AZHYDRO \cite{OSCAR} code,
based on Refs.~\cite{Sollfrank:1990qz,Sollfrank:1991xm}.

\begin{figure}
\vspace{0.6cm}
\begin{center}
\includegraphics[width=0.48\linewidth]{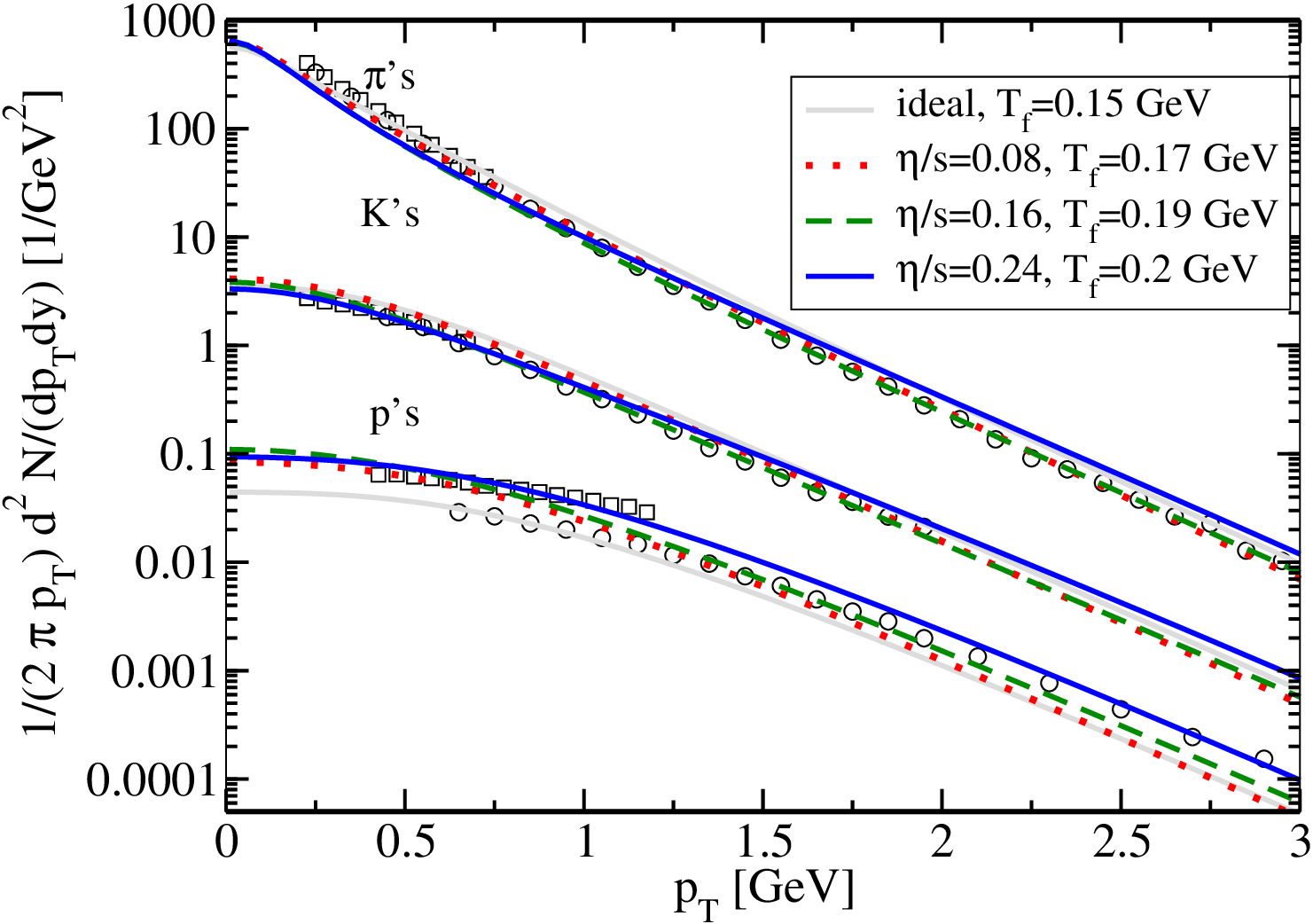}
\hspace*{0.25cm}
\includegraphics[width=0.48\linewidth]{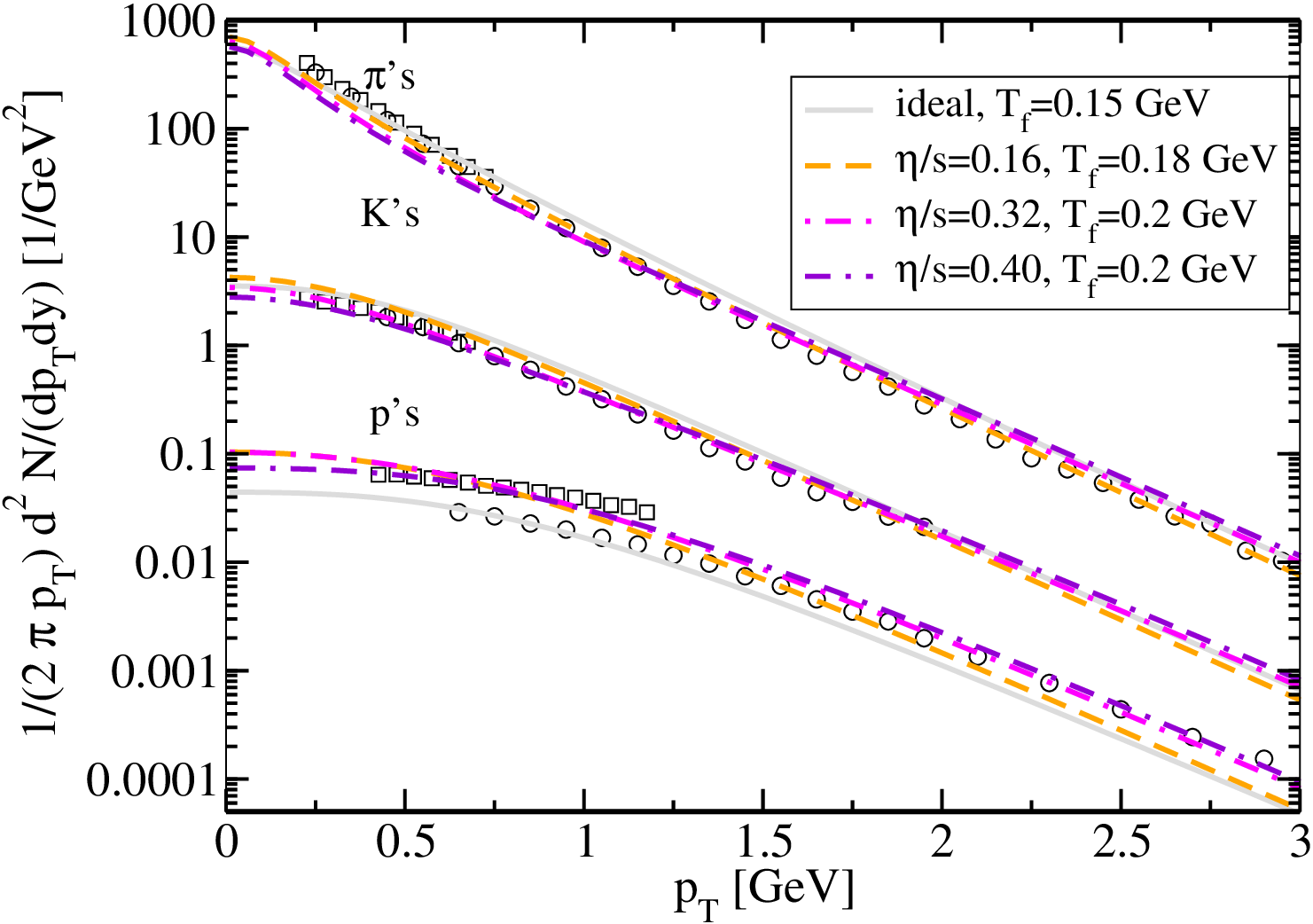}
\end{center}
\caption{Viscous hydrodynamic fits of pion spectra
to experimental data from PHENIX (circles) and 
STAR (squares) \cite{Adler:2003cb,Adams:2003xp},
for relaxation time values $\tau_\Pi=\frac{\eta}{s}\frac{6}{T}$
(left) and $\tau_\Pi=\frac{\eta}{s}\frac{1.5}{T}$ (right).
Also shown are results for kaons and protons, scaled
by $0.1$ and $0.01$, respectively. (The hydrodynamic results
and the STAR data include weak decays while the PHENIX data
does not). 
See text for details.}
\label{fig:specs}
\end{figure}

The initial central temperature $T_0$ and the freeze-out temperature
$T_f$ are adjusted such that both the normalization and slope 
of the resulting pion spectrum is in reasonable agreement with
experimental data \cite{Adler:2003cb,Adams:2003xp}.
This is the procedure adopted in ideal hydrodynamics \cite{Huovinen:2006jp},
and -- as anticipated in Ref.~\cite{Baier:2006um,Baier:2006gy} -- can be carried
through also for non-vanishing $\eta/s$, as is shown in Fig.~\ref{fig:specs}. 
Usually, in addition the initialization time $\tau_0$ 
is allowed to vary in order to obtain a best fit also of the elliptic
flow experimental data.
Since in this work only central collisions are studied, the
choice $\tau_0=1$ fm/c is adopted, but results should not depend
strongly on this choice.

In Fig.~\ref{fig:specs}, results for the spectra of pions, kaons and protons 
for various hydrodynamic runs with different values of shear viscosity
are shown together with experimental data from PHENIX 
\cite{Adler:2003cb} and STAR
\cite{Adams:2003xp}
experiments for the most central $5$\% of Au+Au collisions
at $\sqrt{s_{NN}}=200$ GeV. It should be noted that 
no chemical potential is included in the equation of state,
and therefore a distinction between particles and anti-particles is 
not possible. As a consequence, the spectrum of protons cannot
be expected to match the experimental data, but is included 
in order to demonstrate that it is at least close
to the experimental result.

In general, increasing the value of $\eta/s$ and leaving the initial
and freeze-out conditions unchanged tends to make resulting 
spectra flatter \cite{Teaney,Muronga:2004sf,Chaudhuri:2005ea,Baier:2006gy} 
and hence in some sense mimics increasingly stronger transverse flow.
Starting from initial/freeze-out conditions for which ideal hydrodynamics
fits the experimental spectra and smoothly ``turning on'' viscosity,
one has to counteract the effect of $\eta/s$ by reducing the buildup
of transverse flow to keep the spectra in agreement with 
the experimental data. 
Both decreasing the initial central temperature and increasing
the freeze-out temperature reduces the amount of hydrodynamic transverse
flow at freeze-out.
The former significantly affects the total pion multiplicity
whereas the latter does not, offering a convenient way of
decreasing total transverse flow while keeping the overall 
meson multiplicity close to data.
However, since 
the concept of the Cooper-Frye freeze-out mechanism including 
the effects from resonance decays probably does not make sense
at temperatures that are too high (far above $T_c\simeq 175$ MeV), an
upper limit $T_f\le 200$ MeV is imposed. This limiting 
$T_f$ is much larger than what is typically used in ideal
hydrodynamic model fits. However, in ideal hydrodynamics
the system interactions by definition always keep the system
in perfect thermal equilibrium, while the presence of viscosity
means that interactions are not as efficient, allowing for departures
from equilibrium. As a consequence, an earlier freezing-out and 
thus a higher $T_f$ for viscous hydrodynamics as compared
to ideal hydrodynamics is to be expected.

\begin{table}
\begin{center}
\begin{tabular}{|c|c|c|}
\hline
& $\frac{dN_{\pi, \rm visc}}{dy}/\frac{dN_{\pi, \rm ideal}}{dy}$ &
$\frac{dN_{K, \rm visc}}{dy}/\frac{dN_{K, \rm ideal}}{dy}$ \\
\hline
$\eta/s=0.08$ & 1.06& 1.06\\
\hline
$\eta/s=0.16$, $\tau_\Pi=\frac{\eta}{s}\frac{6}{T}$ & 1.12& 1.12\\
\hline
$\eta/s=0.16$, $\tau_\Pi=\frac{\eta}{s}\frac{1.5}{T}$ & 1.12& 1.12\\
\hline
$\eta/s=0.24$, $\tau_\Pi=\frac{\eta}{s}\frac{6}{T}$ & 1.15& 1.15\\
\hline
$\eta/s=0.24$ $\tau_\Pi=\frac{\eta}{s}\frac{1.5}{T}$ & 1.18& 1.19\\
\hline
$\eta/s=0.32$ & 1.23& 1.23\\
\hline
$\eta/s=0.40$ & 1.28& 1.28\\
\hline
\end{tabular}
\end{center}
\caption{Fraction of multiplicity due to viscous entropy production:
shown are the ratios of total multiplicities for pions and kaons
calculated in viscous and ideal hydrodynamics, for identical
initial/freeze-out conditions.}
\label{tab:one}
\end{table}

In Fig.~\ref{fig:specs}, results are shown for  
a relaxation time value of 
$\tau_\Pi=\frac{\eta}{s}\frac{6}{T}$ (left) and 
$\tau_\Pi=\frac{\eta}{s}\frac{1.5}{T}$ (right) and various $\eta/s$.
For $\tau_\Pi=\frac{\eta}{s}\frac{6}{T}$
the values of $T_0$ used to generate the figure are 
$T_0=0.34,0.33,0.33$ GeV for $\eta/s=0.08,0.16,0.24$.
For $\tau_\Pi=\frac{\eta}{s}\frac{1.5}{T}$
the values are
$T_0=0.34,0.32,0.31$ GeV for $\eta/s=0.16,0.32,0.40$.
From this figure, it can be seen that for the larger value of $\tau_\Pi$, 
the pion (and kaon) spectra in viscous hydrodynamics can be made
to agree reasonably well with experimental data up to $\eta/s\sim 0.25$
by mainly changing the freeze-out temperature $T_f$. At higher values
of $\eta/s$, the pion spectrum becomes either too flat (when leaving
$T_0,T_f$ unchanged), the pion multiplicity becomes too low (when $T_0$ is
lowered such that the slope matches the data) or one has to choose
an unreasonably large $T_f>200$ MeV.

The same is true when adopting a smaller value of $\tau_\Pi$, but
it is possible to have reasonable agreement with experimental data 
up to $\eta/s \sim 0.4$ (see Fig.~\ref{fig:specs}), about twice the bound from 
the larger value of $\tau_\Pi$.
A more detailed study of the effect of
changing $\tau_\Pi$ is currently in progress \cite{prep}.

For non-vanishing viscosity, the hydrodynamic evolution is no
longer isentropic. Some part of the final multiplicity is therefore
due to entropy production; in order to give some quantitative
answers how much entropy is created, it is instructive to consider
the ratio of meson multiplicities at non-vanishing
$\eta/s$ and ideal hydrodynamics for identical initial/freeze-out conditions.
Using the initial/freeze-out conditions for which the viscous
calculations match the experimental pion spectrum, one finds the
results shown in Table \ref{tab:one}. It seems 
plausible to infer from Table \ref{tab:one}
that final multiplicities are increased 
by a fraction of $\sim0.75\, \eta/s$ with respect to ideal hydrodynamics
due to viscous entropy production. Roughly, this translates
to a percentage from $50\, \eta/s$ to $75\, \eta/s$ 
of the final multiplicity being due to 
viscous effects.

\subsection{HBT Radii}

\begin{figure}
\vspace{0.6cm}
\begin{center}
\includegraphics[width=0.48\linewidth]{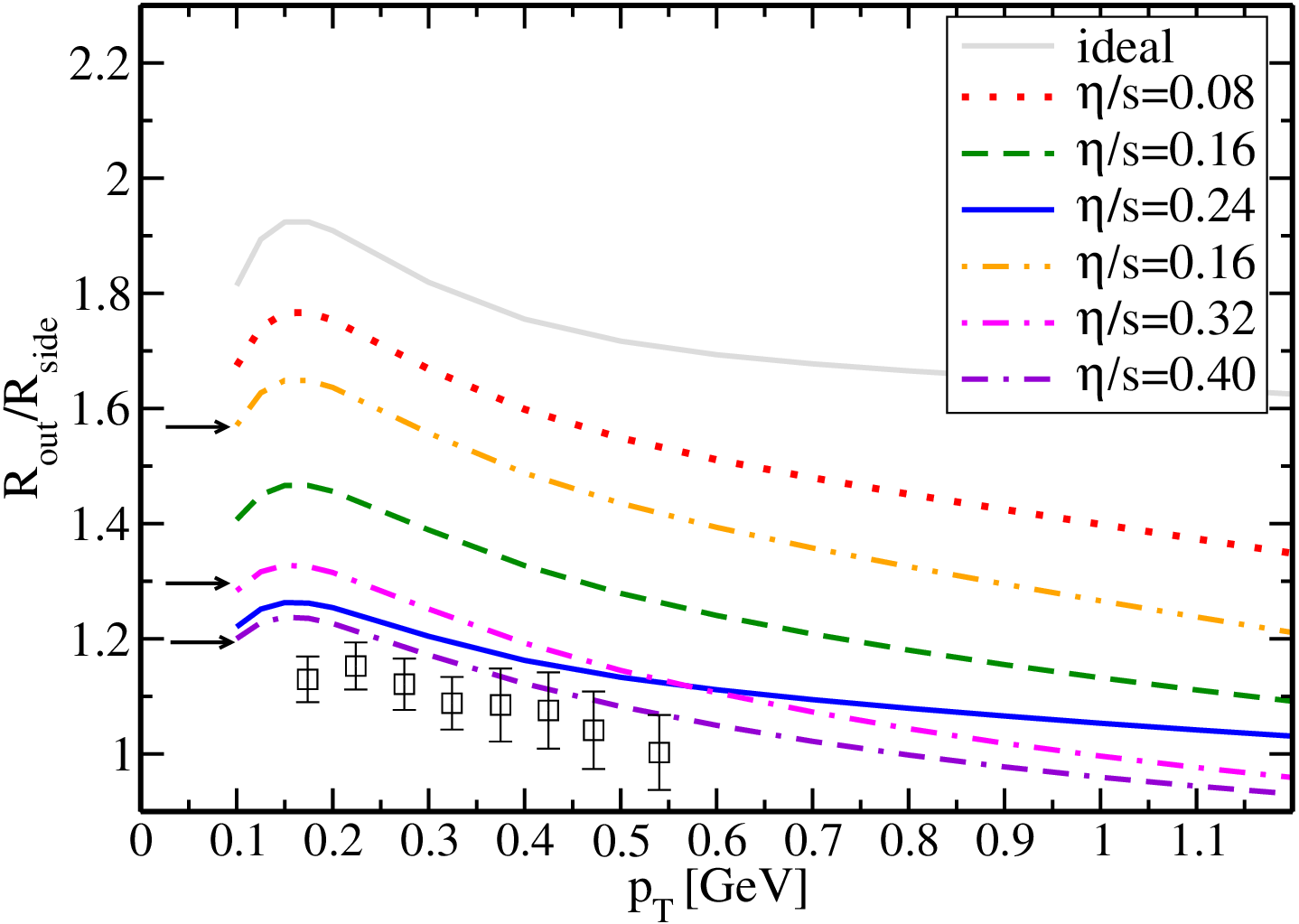}
\hspace*{0.25cm}
\includegraphics[width=0.48\linewidth]{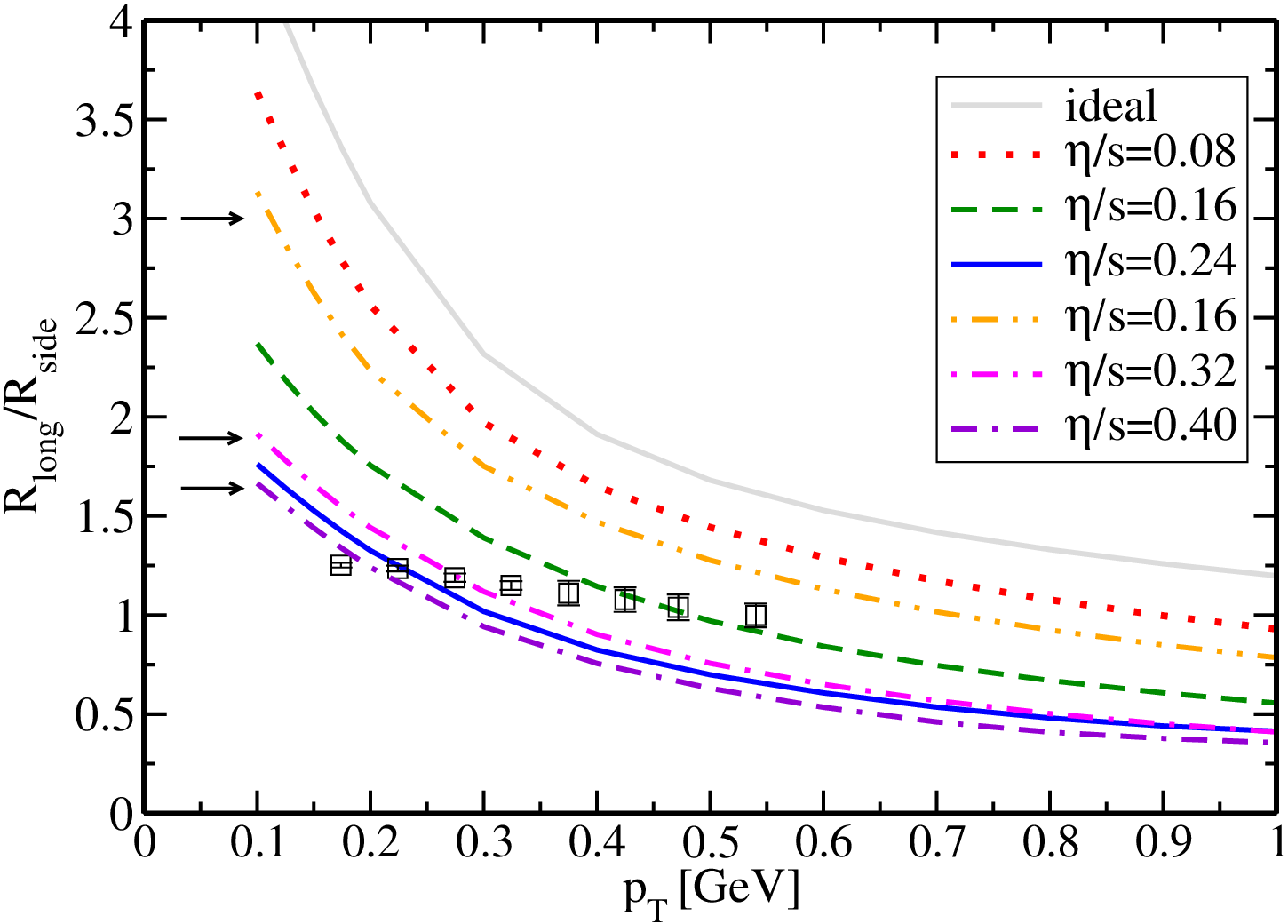}
\end{center}
\caption{Pion HBT Radii from hydrodynamics compared to 
data from the STAR experiment \cite{Adams:2004yc}. Shown are the ratios
$R_{\rm out}/R_{\rm side}$ (left) and $R_{\rm long}/R_{\rm side}$ (right)
for ideal and viscous hydrodynamics (with $\tau_\Pi=\frac{\eta}{s}\frac{6}{T}$
except for results indicated by arrows where 
$\tau_\Pi=\frac{\eta}{s}\frac{1.5}{T}$). See text for details.}
\label{fig:HBTs}
\end{figure}

Using the definition of the HBT radii from the previous section
and the initial/freeze-out conditions for which the viscous hydrodynamic
calculation matches the experimental pion spectrum, one can proceed to
compare the viscous hydrodynamic results for the HBT radii to experimental
data. This is done in Fig.~\ref{fig:HBTs}, where 
the ratios $R_{\rm out}/R_{\rm side}$ and $R_{\rm long}/R_{\rm side}$ 
for various values of $\eta/s$ are compared to data from STAR 
\cite{Adams:2004yc}. The ratio $R_{\rm long}/R_{\rm out}$ is of
special interest since in ideal hydrodynamics it has been notoriously 
hard to obtain results that are not far above 
the experimental data (which sometimes is known as the ``HBT puzzle'').
It has been argued in Refs.~\cite{Dumitru:2002sq,Teaney,Muronga:2004sf}
that non-vanishing viscosity may decrease this ratio and thus 
bring it closer to the data. Fig.~\ref{fig:HBTs} represents
the first result of the ratio $R_{\rm long}/R_{\rm out}$ for 
non-vanishing viscosity and initial/freeze-out conditions for which
the experimental pion spectrum is matched at the same time.

\begin{figure}
\vspace{0.6cm}
\begin{center}
\includegraphics[width=0.48\linewidth]{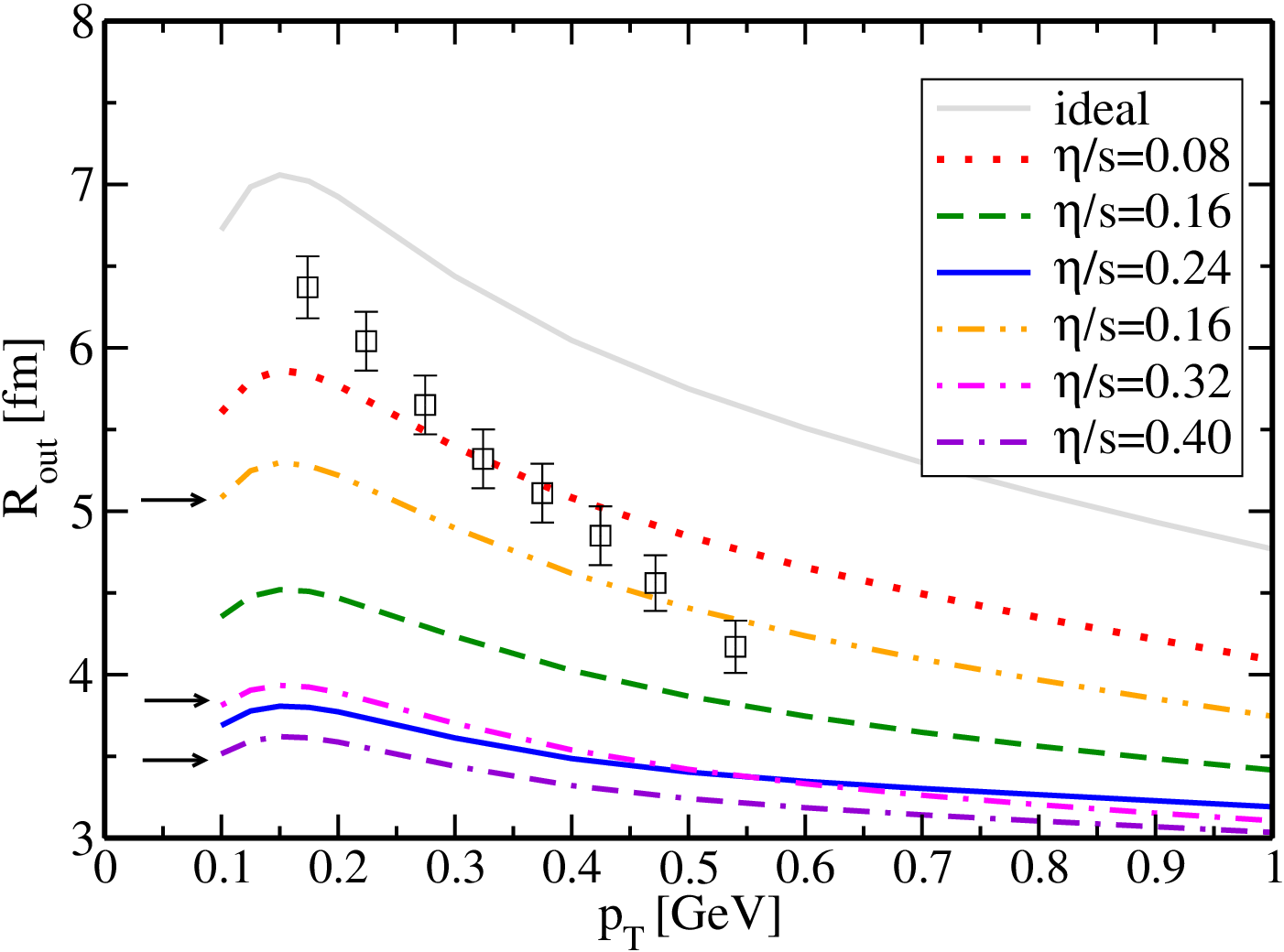}
\hspace*{0.25cm}
\includegraphics[width=0.48\linewidth]{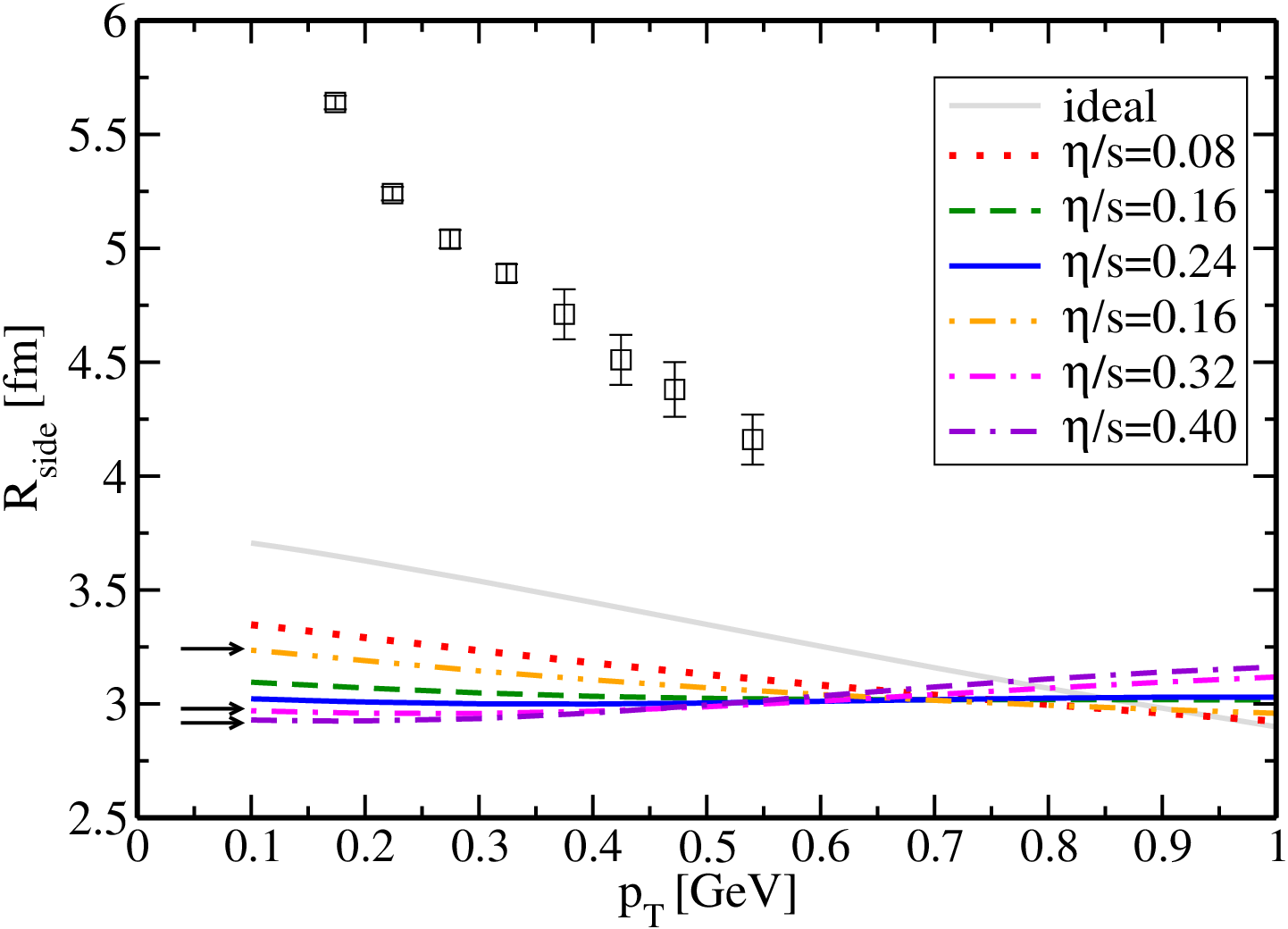}
\end{center}
\caption{Pion HBT Radii from hydrodynamics compared to 
data from the STAR experiment \cite{Adams:2004yc}. Shown are 
$R_{\rm out}$ (left) and $R_{\rm side}$ (right)
for ideal and viscous hydrodynamics (with $\tau_\Pi=\frac{\eta}{s}\frac{6}{T}$
except for results indicated by arrows where 
$\tau_\Pi=\frac{\eta}{s}\frac{1.5}{T}$). See text for details.}
\label{fig:HBT2}
\end{figure}

As can be seen from this figure, the ratio $R_{\rm out}/R_{\rm side}$
approaches the experimental data as viscosity is increased.
Indeed, for all but the lowest $p_\perp$ values the ratio 
is consistent with the data (within error bars) for 
the highest value of viscosity where
the pion spectrum still matches the experimental data.
Remarkable as this may be, it is unlikely that the presence of
viscosity completely solves the HBT puzzle,
the reason being that even though
the ratio $R_{\rm out}/R_{\rm side}$ in the viscous hydrodynamic calculations
moves close to data, the absolute values $R_{\rm out}$, $R_{\rm side}$ 
tend to be below the experimental values, as can be seen from 
Fig.~\ref{fig:HBT2}. Put differently, the agreement of
$R_{\rm out}/R_{\rm side}$ with data is achieved mainly by lowering
$R_{\rm out}$, while $R_{\rm side}$ is hardly affected and always
stays much below the experimental values.

At present, it is not excluded that a match of the viscous hydrodynamic
$R_{\rm side}$ with experimental data can be achieved
when e.g. when changing also the hydrodynamic initialization time $\tau_0$,
or the initial energy density profile \cite{Huovinen:2006jp}.
However, the mismatch of $R_{\rm side}$ with data
could also indicate that other effects, such as the replacement
of the sharp hypersurface freeze-out by a more realistic continuous
emission of particles \cite{Grassi:2000ke,Sinyukov:2002if} 
have to be taken into account
to obtain a proper description of the experimental HBT radii.

Also shown in Fig.~\ref{fig:HBTs} is the ratio $R_{\rm long}/R_{\rm side}$,
compared to experimental data \cite{Adams:2004yc}. In this case,
the viscous hydrodynamic result can be made to intersect the
experimental results, but the too steep slope seems to prevent a nice match
to data. Again the absolute values tend to be below the experimental data,
also perhaps indicating the need for a more realistic freeze-out
treatment.

\section{Conclusions}
\label{sec:five}

Using a simple numerical code to solve the causal viscous
hydrodynamic equations for the case of central heavy-ion collisions, 
it has been shown that both pion total multiplicities and 
spectral slopes for Au+Au collisions at $\sqrt{s_{NN}}=200$ GeV 
can be matched to experimental data 
for moderate viscosities\footnote{The C++ code
including results is available from http://hep.itp.tuwien.ac.at/\~\,paulrom/}. 
The bound on $\eta/s$ up to which this matching is possible depends on the
value of the relaxation time (which is a priori a free parameter of
the causal viscous hydrodynamic framework) and on the hydrodynamic
initialization time $\tau_0$. 
Setting $\tau_0=1$ fm/c, one obtains $\eta/s\lesssim 0.25$ 
for a value of the relaxation time $\tau_\Pi$ 
that corresponds to the weak-coupling
QCD result, and $\eta/s\lesssim 0.4$ for a somewhat
smaller value of $\tau_\Pi$.
For all the viscous hydrodynamic calculations that result in
a pion spectrum matching the experimental data, it was found
that roughly $50\, \eta/s$ to $75\, \eta/s$ percent 
of the final multiplicity
is due to viscous entropy production.

For hydrodynamic parameters that allow to 
match the experimental pion spectrum,
it was shown that the ratio of HBT radii $R_{\rm out}/R_{\rm side}$
approaches and eventually matches 
the experimental result as $\eta/s$ is increased. 
The absolute values of $R_{\rm out}$, $R_{\rm side}$, however,
tend to be much below the experimental values, suggesting
that a more elaborate freeze-out treatment (or other
effects, see e.g. Ref.~\cite{Cramer:2004ih}) 
might be necessary to achieve agreement.

The above bound on the ratio $\eta/s\lesssim 0.4$ is interesting
since it lies roughly half-way in between the weak-coupling
QCD result \cite{Trans1,Trans2} and the strong coupling
${\mathcal N}=4$ SYM result \cite{Kovtun:2004de,Janik:2006ft}.
If this bound was saturated, this could suggest that
the quark-gluon plasma created at the highest RHIC energies
is neither a very weakly nor a very strongly coupled plasma, but
rather a border-case between these two. While this may not be
the most elegant of possibilities that nature could have chosen, 
a not-weakly, not-strongly coupled plasma which behaves
as a non-ideal fluid might be the most
realistic assessment of the current status of RHIC data
(see Ref.~\cite{Romatschke:2006bb,Huot:2006ys,Blaizot:2006tk} 
for related conclusions).

To decide, and maybe extract a value of the ratio
$\eta/s$ (and at the same time $\tau_\Pi$) from RHIC data,
it seems necessary to extend this work to the treatment of
non-central collisions, and thus viscous hydrodynamic results
for elliptic flow. 

\acknowledgments

I would like to thank R.~Baier, D.~d'Enterria, U.~Heinz, P.~Huovinen, M.~Laine,
G.A.~Miller and S.~Salur for fruitful discussions. Special thanks go to
P.~Huovinen for providing me with the data file for the resonance
decays and to M.~Laine for providing the tabulated equation of state.
This work was supported by the US Department of Energy, grant
number DE-FG02-00ER41132.


\begin{thebibliography}{99} 
\bibitem{Hisc}
W.A.~Hiscock and L.~Lindblom, Phys.\ Rev.\ D {\bf 31}, 725 (1985).
\bibitem{IS}
W.~Israel, Ann.Phys. {\bf 100} (1976) 310;
W.~Israel and J.M.~Stewart, Phys. Lett. {\bf 58A} (1976)  213;
W.~Israel and J.M.~Stewart, Ann.Phys. {\bf 118}, (1979) 341.
\bibitem{LMR}
I-Shih~Liu, I.~M\"uller and T.~Ruggeri, Ann.Phys. {\bf 169} (1986) 191.

\bibitem{Muronga:2001zk}
  A.~Muronga,
  Phys.\ Rev.\ Lett.\  {\bf 88} (2002) 062302
  [Erratum-ibid.\  {\bf 89} (2002) 159901].
\bibitem{Muronga:2003ta}
  A.~Muronga,
  Phys.\ Rev.\ C {\bf 69} (2004) 034903.
\bibitem{Baier:2006um}
  R.~Baier, P.~Romatschke and U.~A.~Wiedemann,
  Phys.\ Rev.\ C {\bf 73} (2006) 064903.

\bibitem{Koide:2006ef}
  T.~Koide, G.~S.~Denicol, Ph.~Mota and T.~Kodama,
  arXiv:hep-ph/0609117.

\bibitem{Muronga:2004sf}
  A.~Muronga and D.~H.~Rischke,
  arXiv:nucl-th/0407114.
%

\bibitem{Chaudhuri:2005ea}
  A.~K.~Chaudhuri and U.~W.~Heinz,
  J.\ Phys.\ Conf.\ Ser.\  {\bf 50} (2006) 251.

\bibitem{Baier:2006gy}
  R.~Baier and P.~Romatschke,
  arXiv:nucl-th/0610108.

\bibitem{Trans1}
  P.~Arnold, G.~D.~Moore and L.~G.~Yaffe,
  JHEP {\bf 0011} (2000) 001.

\bibitem{Trans2}
  P.~Arnold, G.~D.~Moore and L.~G.~Yaffe,
  JHEP {\bf 0305} (2003) 051.
\bibitem{Kovtun:2004de}
  P.~Kovtun, D.~T.~Son and A.~O.~Starinets,
  Phys.\ Rev.\ Lett.\  {\bf 94} (2005) 111601.
%
\bibitem{Janik:2006ft}
  R.~A.~Janik,
  Phys.\ Rev.\ Lett.\  {\bf 98} (2007) 022302.

\bibitem{Asakawa:2006tc}
  M.~Asakawa, S.~A.~Bass and B.~Muller,
  Phys.\ Rev.\ Lett.\  {\bf 96} (2006) 252301.

\bibitem{Mrowczynski:2005ki}
  S.~Mrowczynski,
  Acta Phys.\ Polon.\ B {\bf 37} (2006) 427.


\bibitem{Arnold:2006fz}
  P.~Arnold, C.~Dogan and G.~D.~Moore,
  Phys.\ Rev.\ D {\bf 74} (2006) 085021.


\bibitem{Heinz:2005bw}
  U.~W.~Heinz, H.~Song and A.~K.~Chaudhuri,
  Phys.\ Rev.\ C {\bf 73} (2006) 034904.

\bibitem{Muronga:2006zw}
  A.~Muronga,
  arXiv:nucl-th/0611090.

\bibitem{prep}
R.~Baier, P.~Romatschke and A.~Starinets, in preparation.


\bibitem{Kolb:2001qz}
  P.~F.~Kolb, U.~W.~Heinz, P.~Huovinen, K.~J.~Eskola and K.~Tuominen,
  Nucl.\ Phys.\ A {\bf 696} (2001) 197.

\bibitem{fodor}
Y.~Aoki, Z.~Fodor, G.~Endrodi, S.D.~Katz and K.K.~Szabo,
Nature {\bf 443} (2006) 675.

\bibitem{Laine:2006cp}
  M.~Laine and Y.~Schroder,
  Phys.\ Rev.\ D {\bf 73} (2006) 085009.



\bibitem{CooperFrye}
F.~Cooper and G.~Frye, Phys.\ Rev.\ D {\bf 10}, 186 (1974).

\bibitem{Rischke:1996em}
  D.~H.~Rischke and M.~Gyulassy,
  Nucl.\ Phys.\ A {\bf 608} (1996) 479.

\bibitem{Teaney}
  D.~Teaney,
  Phys.\ Rev.\ C {\bf 68} (2003) 034913.
  %


\bibitem{Schlei:1992jj}
  B.~R.~Schlei, U.~Ornik, M.~Plumer and R.~M.~Weiner,
  Phys.\ Lett.\ B {\bf 293} (1992) 275.

\bibitem{Pratt:1986ev}
  S.~Pratt,
  Phys.\ Rev.\ D {\bf 33} (1986) 72.

\bibitem{Bertsch:1989vn}
  G.~F.~Bertsch,
  Nucl.\ Phys.\ A {\bf 498} (1989) 173C.

\bibitem{OSCAR}
Version 0.2, available from
http://nt3.phys.columbia.edu/people/molnard/OSCAR/

\bibitem{Sollfrank:1990qz}
  J.~Sollfrank, P.~Koch and U.~W.~Heinz,
  Phys.\ Lett.\ B {\bf 252} (1990) 256.

\bibitem{Sollfrank:1991xm}
  J.~Sollfrank, P.~Koch and U.~W.~Heinz,
  Z.\ Phys.\ C {\bf 52} (1991) 593.

\bibitem{Adler:2003cb}
  S.~S.~Adler {\it et al.}  [PHENIX Collaboration],
  Phys.\ Rev.\ C {\bf 69} (2004) 034909.

\bibitem{Adams:2003xp}
  J.~Adams {\it et al.}  [STAR Collaboration],
  Phys.\ Rev.\ Lett.\  {\bf 92} (2004) 112301.

\bibitem{Huovinen:2006jp}
  P.~Huovinen and P.~V.~Ruuskanen,
  arXiv:nucl-th/0605008.

\bibitem{Adams:2004yc}
  J.~Adams {\it et al.}  [STAR Collaboration],
  Phys.\ Rev.\ C {\bf 71} (2005) 044906.

\bibitem{Dumitru:2002sq}
  A.~Dumitru,
  arXiv:nucl-th/0206011.

\bibitem{Grassi:2000ke}
  F.~Grassi, Y.~Hama, S.~S.~Padula and O.~J.~Socolowski,
  Phys.\ Rev.\ C {\bf 62} (2000) 044904.

\bibitem{Sinyukov:2002if}
  Yu.~M.~Sinyukov, S.~V.~Akkelin and Y.~Hama,
  Phys.\ Rev.\ Lett.\  {\bf 89} (2002) 052301.

\bibitem{Cramer:2004ih}
  J.~G.~Cramer, G.~A.~Miller, J.~M.~S.~Wu and J.~H.~S.~Yoon,
  Phys.\ Rev.\ Lett.\  {\bf 94} (2005) 102302.

\bibitem{Romatschke:2006bb}
  P.~Romatschke,
Phys.\ Rev.\ C {\bf 75}, (2007) 014901. 

\bibitem{Huot:2006ys}
  S.~C.~Huot, S.~Jeon and G.~D.~Moore,
  arXiv:hep-ph/0608062.

\bibitem{Blaizot:2006tk}
  J.~P.~Blaizot, E.~Iancu, U.~Kraemmer and A.~Rebhan,
  arXiv:hep-ph/0611393.

\end{thebibliography}
\end{document}